\documentclass[aps,nopacs,nofootinbib]{revtex4}
\usepackage{epsf,latexsym}

\begin{document}

\title{Null-null components of the generalized Einstein tensor for Lovelock models}
\author{Alessandro Pesci}
\email{pesci@bo.infn.it}
\affiliation
{INFN-Bologna, Via Irnerio 46, I-40126 Bologna, Italy}

\begin{abstract}
For spherical symmetry,
we provide expressions for the radial null-null components
of the generalized Einstein tensor $E_{ab}$ for Lovelock models
for diagonal $E_{ab}$
in terms of the metric and
of the radial null-null components of the Ricci tensor.
We show they can be usefully employed for example 
in obtaining 
the Birkhoff-like theorem for Lovelock models,
as well as in providing a relation
between the null energy and null convergence
conditions for the same models.
\end{abstract}


\maketitle

$ $
\section{Introduction}

Calculations in spherically symmetric spacetimes
are far easier if the metric of the spacetime under consideration
has time-time component and $r-r$ component ($r$ is areal radius) 
in the form $g_{tt} g_{rr} =-1$, like the Schwarzschild solution
to Einstein's equations.
When one deals 
with some metric theory of gravity,
and starts cosidering first the case
with as much as symmetry and simplicity as possible, 
one would like then for example to know 
whether the metric of the spherically symmetric vacuum solutions 
of the theory is necessarily of this form or not.
A general condition for a static spherically symmetric metric to have
this form has been given in \cite{JacF}, in terms of the vanishing
of the radial null-null components of the Ricci tensor.
This condition can equivalently be thought, of course,
as the vanishing of the radial null-null components of the Einstein tensor.

For spherically symmetric vacuum solutions to Einstein's equations,
one knows thus
that the (static) metric does have the form above;
this can also be viewed as 
a manifestation of Birkhoff's theorem \cite{Jeb, Bir} at work.
When one goes, however, to general theories of gravity,
what drives the motion is no longer the Einstein tensor
but the generalized Einstein tensor.
For vacuum solutions,
what one knows is thus the vanishing 
of the radial null-null components of the generalized Einstein tensor
(and of the tensor itself, of course),
not of these same components for Einstein or Ricci.
In this context,
the point at issue apparently is 
to know when
the vanishing of those components for generalized Einstein
means their vanishing for Ricci.

Here, we investigate this for Lovelock models.
From the Birkhoff-like theorem for 
generic such models \cite{CD, Zeg, DesA},
one already knows that the metric for spherically symmetric vacuum
solutions must have the form above,
and thus that for these models
the vanishing of the radial null-null components for generalized Einstein
must imply their vanishing for Ricci. 
One interesting thing, however,
could be
to read this directly from the expression of
these components for the generalized Einstein tensor.
Our aim is to provide a multi-purpose expression 
for these components in terms of the metric,
and of the same components for Ricci.
From the generic vanishing of it
one should directly read
the vanishing of the radial null-null components of Ricci,
thus obtaining the Birkhoff-like theorem.
Another use of this expression, 
is in exploring the relation
between the null energy condition (NEC) and 
the null convergence condition (NCC) in the radial direction
for Lovelock models.
%

\section{Statement of the question}

In $D$-dimensional spacetime, 
we consider
gravitational Lagrangians $L$ with 
$L = L(g^{ab}, {R^a}_{bcd})$, being $g^{ab}$ the metric
and ${R^a}_{bcd}$ the Riemann tensor 
(latin labels span all the $D$ coordinates, $\{x^a\}$),
that is with general dependence on metric and Riemann tensor
but with no dependence on derivatives
of the latter.
The equations of motion for the field $g^{ab}$ we get
for an action $I$ with variation
$
\delta I = 
\int d^D x \ \delta (\sqrt{-g} L)
- \frac{1}{2} \ \int d^D x \ \sqrt{-g} \ T_{ab} \ \delta g^{ab},
$
are 
\begin{eqnarray}\label{fie}
2 \ E_{ab} = T_{ab},
\end{eqnarray}
where $T_{ab}$ is the energy-momentum tensor
and

\begin{eqnarray}\label{Eab}
E_{ab} =
U_{ab} - 2 \ \nabla^i\nabla^j P_{aijb}
\end{eqnarray}
is the generalized Einstein tensor,
with
$
U_{ab} =
\frac{1}{\sqrt{-g}} \frac{\partial(\sqrt{-g} L)}{\partial g^{ab}}
$
and
$
{P_a}^{ijb} = \frac{\partial L}{\partial {R^a}_{ijb}}
$
(see e.g. \cite{PadM}).
$U_{ab}$ and $P_{abcd}$ have the same symmetries 
in their indices
as $g_{ab}$ and $R_{abcd}$ respectively.

Within this set of Lagrangians,
we consider the subset of
Lovelock Lagrangians (LL)
\cite{LanA,LanB,Lov},
i.e. linear combinations
of a suitable, finite number
of pure Lovelock terms, 
each one writable as

\begin{eqnarray}\label{LLL}
L_{(m)} \equiv
\frac{1}{2^m} \ \delta^{c_1 d_1 c_2 d_2 ... c_m d_m}_{a_1 b_1 a_2 b_2 ... a_m b_m}
\ R^{a_1 b_1}_{c_1 d_1} R^{a_2 b_2}_{c_2 d_2} ... R^{a_m b_m}_{c_m d_m} 
= \frac{1}{m} \ P^{cd}_{ab} R^{ab}_{cd},
\end{eqnarray}
where $m = 1, 2, ...$ is the order of the term.
Here, $\delta^{c_1 d_1 c_2 d_2 ... c_m d_m}_{a_1 b_1 a_2 b_2 ... a_m b_m}$ 
are $D$-dimensional ``permutation tensors'' \cite{MisC}
of rank $2 m$,
and the equality comes either
differentiating directly the expression for $L(m)$
or making use of Euler's theorem, 
being $L_{(m)}$ a homogeneous function of degree $m$ in $R^{ab}_{cd}$.
$
L_{(1)} \equiv
\frac{1}{2} \ \delta^{cd}_{ab} \ R^{ab}_{cd} =
R
$  
is the Einstein-Hilbert Lagrangian,
given by the scalar curvature $R$.
$
L_{(2)} \equiv
\frac{1}{4} \ \delta^{c_1 d_1 c_2 d_2}_{a_1 b_1 a_2 b_2} 
\ R^{a_1 b_1}_{c_1 d_1} R^{a_2 b_2}_{c_2 d_2} =
R^{abcd} R_{abcd} - 4 \ R^{ab} R_{ab} + R^2
$
is the Gauss-Bonnet term ($R_{ab}$ is the Ricci tensor).

For LL,

\begin{eqnarray} \label{generic_L}
L = \sum_{m=1}^{[\frac{D-1}{2}]} c_m L_{(m)},
\end{eqnarray} 
with 
$c_m$ real constants. 
Denoting with
$U_{ab(m)}$, $P{{_a}^{ijb}}_{(m)}$ and $E_{ab(m)}$
the tensors in eq. (\ref{Eab})
corresponding to $L_{(m)}$,
in the equations of motion (\ref{fie}) we have
$E_{ab} = \sum_{m=1}^{[\frac{D-1}{2}]} c_m E_{ab(m)}$,
with

\begin{eqnarray}\label{Eabm}
E_{ab(m)} &=&  U_{ab(m)} - 2 \ \nabla^i\nabla^j P_{aijb(m)}  
= U_{ab(m)} 
= - \frac{1}{2} \ g_{ab} L_{(m)} + 
\frac{\partial L_{(m)}}{\partial g^{ab}} \nonumber \\ 
&=& - \frac{1}{2} \ g_{ab} L_{(m)} + \frac{\partial L_{(m)}}{\partial R^{kl}_{ij}}
\frac{\partial R^{kl}_{ij}}{\partial g^{ab}}
= - \frac{1}{2} \ g_{ab} L_{(m)} + P^{ij}_{kl(m)} 
\ \frac{1}{2} 
\big(\delta^l_a {R^k}_{bij} + \delta^l_b {R^k}_{aij}\big) \nonumber \\
&=& - \frac{1}{2} \ g_{ab} L_{(m)} + 
\frac{1}{2} \big(P^{ij}_{ka(m)} {R^k}_{bij} + P^{ij}_{kb(m)} {R^k}_{aij}\big)
= - \frac{1}{2} \ g_{ab} L_{(m)} + P^{ij}_{kb(m)} {R^k}_{aij},
\end{eqnarray}
where
the last equality follows
from turning out the two terms 
in round brackets of l.h.s. equal \cite{PadO}.
These expressions are only second order in the derivatives of the metric. 
Essential to this, is
the second equality. It is obtained
thanks to the crucial property of these Lagrangians 
of having ${P^{abcd}}_{(m)}$
with zero divergence on each of the indices:

\begin{eqnarray}\label{div}
\nabla_c {P^{abcd}}_{(m)} =
0,
\end{eqnarray} 
and the same for any other index \cite{comment4}.
For $m = 1$,
we get

\begin{eqnarray}\label{GR}
E_{ab(1)} =
- \frac{1}{2} \ g_{ab} L_{(1)} + P^{ij}_{kb(1)} {R^k}_{aij} =
- \frac{1}{2} \ g_{ab} R + \frac{1}{2}\delta^{ij}_{kb} {R^k}_{aij} =
- \frac{1}{2} \ g_{ab} R + R_{ab},
\end{eqnarray}
so that $E_{ab(1)}$ is the Einstein tensor,
and the equations of motion
$
2 E_{ab(1)} =
T_{ab}
$
are the Einstein equations (in Planck units, 
and without a $1/16\pi$ factor in the l.h.s.). 
For the pure Gauss-Bonnet term,
we get

\begin{eqnarray}
E_{ab(2)} 
&=& - \frac{1}{2} \ g_{ab} L_{(2)} + P^{ij}_{kb(2)} {R^k}_{aij} 
= - \frac{1}{2} \ g_{ab} L_{(2)} + 
\frac{1}{2} \ \delta^{ij c_2 d_2}_{kb a_2 b_2} 
\ R^{a_2 b_2}_{c_2 d_2} {R^k}_{aij}  \nonumber \\ 
&=& - \frac{1}{2} \ g_{ab} L_{(2)} +
2\ {R^k}_{aij} R^{ij}_{kb} + 4\ R^j_k {R^k}_{abj} +
2\ R_{ab} R - 4\ R^j_a R_{bj}.
\end{eqnarray}

The Lagrangians are chosen with cosmological constant $\Lambda = 0$. 
The effects of any additional term $\Lambda g_{ab}$ with $\Lambda \ne 0$
in the Lagrangian,  
can conveniently be described
leaving the Lagrangian as it is, i.e. without cosmological term,
and introducing, among the sources, a cosmological ideal fluid
with stress-energy tensor $(T_\Lambda)_{ab} = -2 \ \Lambda g_{ab}$.  

Considering any null field $n^a$ in vacuum (vac) or
in vacuum with cosmological constant (cosmovac),
we have 
$T_{ab} \ n^a n^b = 0$,
and thus
$E_{ab} \ n^a n^b =0$.
Considering, in particular, some piece of spacetime with spherical symmetry
with $l^a$ denoting a generic radial null vector field,
this means 
$E_{ab} \ l^a l^b = 0$.

When $m=1$,
from (\ref{GR}) we see that 
$E_{ab(1)} \ l^a l^b = 0$
is manifestly equivalent to
$R_{ab} \ l^a l^b = 0$.
In the $m=2$ case, and even more so for $m>2$, 
the dependence on $R^{ab}_{cd}$ in $E_{ab(m)}$
is more involved,
so that
the just mentioned equivalence is not manifest, 
if present at all 
(but we know this must actually be somehow the case
from Birkhoff-like theorem for Lovelock models).
Our aim is to try to work out a general expression for
$E_{ab} \ l^a l^b$  in terms of the components of the metric
for Lovelock models
for diagonal $E_{ab}$,
somehow generalizing what is done 
in certain derivations 
(in \cite{PadM}, for instance)
of Birkhoff's theorem
in general relativity.
The idea/hope is that
this expression can be put also in a form such that
$E_{ab} \ l^a l^b = 0$
turns out to be,
at least at certain conditions to be investigated,
manifestly equivalent to
$R_{ab} \ l^a l^b = 0$,
and that, among other possible uses,
it can be exploited to derive 
the Birkhoff-like theorem, 
as well as to explore the relation between the
NEC and the NCC in the radial direction
for Lovelock models.

\section{Calculation}

We are going to provide an expression for $E_{ab} \ l^a l^b$
in terms of the metric
for spherically symmetric configurations.
This expression 
turns out to be the same
in the static and non-static cases,
provided $E_{ab}$ 
is diagonal in spherical coordinates
(for a static configuration, this is the case already;
for a non-static configuration, 
the meaning of this is to require $E_{tr} = E_{rt} = 0$).   
Let us consider first the static case.
The general metric for a spherically symmetric, static spacetime 
can be written as

\begin{eqnarray}\label{metric}
ds^2 
= -A \ dt^2 + B \ dr^2 + r^2 \ d\Omega^2 
= -A \ dt^2 + B \ dr^2 + r^2 \ \sum_{I} \ h_{II}(x) \ (dx^I)^2
\end{eqnarray}
where $A=A(r)$ and $B=B(r)$,
being $r$ the areal radius,  
and where the $x^I$, $I = 3, 4, ... , D$, are chosen to be
the usual angular coordinates 
parametrising the $(D-2)$-dimensional manifold orthogonal
to $(t,r)$ in diagonal form
($
d\Omega^2 =
d\alpha_1^2 + \sin^2\alpha_1  
(d\alpha_2^2 + \sin^2\alpha_2  
(d\alpha_3^2 + \sin^2\alpha_3  
( ... )
$)),
with
$(\alpha_1, ... , \alpha_{D-2}) = x^I$)
and the sum over $I$ is explicitly indicated.
What we have to do,
is to find an expression for $E_{ab(m)} \ l^a l^b$
in terms of $r$, $A$ and $B$.

From (\ref{Eabm}) 
and from the definition of $P^{cd}_{ab(m)}$
we get
\begin{eqnarray}
E_{tt(m)}
&=& 
P^{ij}_{kt(m)} {R^k}_{tij} - \frac{1}{2} g_{tt} L_{(m)} \nonumber \\
&=&
\frac{1}{2^m} \ g_{tt} \
\Big( m \
\delta^{i j c_2 d_2 ... c_m d_m}_{k t a_2 b_2 ... a_m b_m}
\ R^{kt}_{ij} R^{a_2 b_2}_{c_2 d_2} ... R^{a_m b_m}_{c_m d_m}
-\frac{1}{2}
\delta^{c_1 d_1 c_2 d_2 ... c_m d_m}_{a_1 b_1 a_2 b_2 ... a_m b_m}
\ R^{a_1 b_1}_{c_1 d_1} R^{a_2 b_2}_{c_2 d_2} ... R^{a_m b_m}_{c_m d_m} 
\Big) \nonumber \\
&=&
\frac{1}{2^m} \ g_{tt} \
\Big(
-\frac{1}{2}
\delta^{c_1 d_1 c_2 d_2 ... c_m d_m ({\not t})}_{a_1 b_1 a_2 b_2 ... a_m b_m ({\not t})}
\ R^{a_1 b_1}_{c_1 d_1} R^{a_2 b_2}_{c_2 d_2} ... R^{a_m b_m}_{c_m d_m} 
\Big),
\end{eqnarray}
with $\not i$ meaning the index $i$ cannot be present in the string.
For the metric (\ref{metric}),
the only non-vanishing components of the Riemann tensor
are  
$R^{t \alpha}_{t \alpha}$, $R^{r \alpha}_{r \alpha}$
and 
$R^{\alpha\beta}_{\alpha\beta}$ 
$(\alpha \ne \beta)$
(and  those related to these by symmetries), 
and they do not depend on $\alpha$ and $\beta$.
Here and in what follows,
Greek indices denote specific angular components 
and no convention on sum of repeated indices
is assumed for them.
From this and the symmetries of Riemann, we get

\begin{eqnarray}\label{p250}
E_{tt(m)} 
&=&
\frac{1}{2} \ g_{tt} \ \frac{1}{2^m} \
\Big(
- m \ 2^{m+1}
\sum_{\alpha, \alpha_2, \beta_2, ..., \alpha_m, \beta_m \ne}
\ R^{r \alpha}_{r \alpha} R^{\alpha_2 \beta_2}_{\alpha_2 \beta_2} ... R^{\alpha_m \beta_m}_{\alpha_m \beta_m}
- 2^m 
\sum_{\alpha, \beta, \alpha_2, \beta_2, ..., \alpha_m, \beta_m \ne}
\ R^{\alpha \beta}_{\alpha \beta} R^{\alpha_2 \beta_2}_{\alpha_2 \beta_2} ... R^{\alpha_m \beta_m}_{\alpha_m \beta_m}
\Big)
\nonumber \\  
&=&
\frac{1}{2} \ g_{tt} \
\Big[
- m \ 2 (D-2) (D-3) ... (D-2m) \ R^{r \alpha}_{r \alpha} \big(R^{\alpha \beta}_{\alpha \beta}\big)^{m-1}
- (D-2) (D-3) ... (D-2m-1) \big(R^{\alpha \beta}_{\alpha \beta}\big)^m
\Big] 
\nonumber \\
&=&
\frac{1}{2} \ g_{tt} \
(D-2)! \
\frac{1}{(D-2m-1)!} \ \big(R^{\alpha \beta}_{\alpha \beta}\big)^{m-1} \
\Big[
- 2 m \ R^{r \alpha}_{r \alpha} - (D-2m-1) \ R^{\alpha \beta}_{\alpha \beta}
\Big]. 
\end{eqnarray}
The `$\ne$' symbol in the sums means that the sums are taken
with all indices different.
Analogous calculations give

\begin{eqnarray}\label{p253m}
E_{rr(m)}  =
\frac{1}{2} \ g_{rr} \
(D-2)! \
\frac{1}{(D-2m-1)!} \ \big(R^{\alpha \beta}_{\alpha \beta}\big)^{m-1} \
\Big[
- 2 m \ R^{t \alpha}_{t \alpha} - (D-2m-1) \ R^{\alpha \beta}_{\alpha \beta}
\Big] 
\end{eqnarray}
and
$E_{rt(m)} = 0$.

Defining

\begin{eqnarray}\label{polQ}
Q[X] :=
\sum_{m=1}^{[\frac{D-1}{2}]} \frac{m \ c_m}{(D-2m-1)!} \ X^{m-1}
\end{eqnarray}
and
\begin{eqnarray}\label{polW}
W[X] :=
\sum_{m=1}^{[\frac{D-1}{2}]} \frac{c_m}{(D-2m-1)!} \ X^m
=
\int_{0}^{X} Q[{\tilde X}] \ d{\tilde X},
\end{eqnarray}
$\frac{dW}{dX} = Q$,
polynomials in

\begin{eqnarray}\label{X}
X \ = \
R^{\alpha \beta}_{\alpha \beta} (\alpha \ne \beta) \
= \ \frac{1}{r^2} \ \Big(1 - \frac{1}{B}\Big),
\end{eqnarray}
equations (\ref{p250}) and (\ref{p253m}) give

\begin{eqnarray}\label{p251}
E_{tt} =
\frac{1}{2} \ g_{tt} \
(D-2)! \
\Big\{ 
2 \ \big(- R^{r \alpha}_{r \alpha} + X\big) \ Q[X] - (D-1) \ W[X] 
\Big\}
\end{eqnarray}
\begin{eqnarray}\label{p253}
E_{rr} =
\frac{1}{2} \ g_{rr} \
(D-2)! \
\Big\{ 
2 \ \big(- R^{t \alpha}_{t \alpha} + X\big) \ Q[X] - (D-1) \ W[X] 
\Big\}.
\end{eqnarray}

From these and $E_{rt} = 0$, we get

\begin{eqnarray}\label{p240}
E_{ab} \ l^a l^b 
&=&
\mu^2 \ (D-2)! \ Q[X] \ 
\Big[
- g_{tt} g_{rr} \ \big(R^{r \alpha}_{r \alpha}- R^{t \alpha}_{t \alpha}\big)
\Big]
\nonumber \\
&=&
\mu^2 \ (D-2)! \ Q[X] \ \frac{1}{2rB} (AB)^\prime,
\end{eqnarray}
where $l^a$, 
with

\begin{eqnarray}\label{radialnull}
l^t = \mu \ \sqrt{g_{rr}}, \ 
l^r =\pm \mu \ \sqrt{-g_{tt}}, \ 
l^\alpha = 0,
\end{eqnarray}
$\mu$ a function,
is generic radial null.
In the derivation of (\ref{p240}),
use has been made of 
the explicit expressions
$g_{tt} = -A$,
$g_{rr} = B$, 
$
R^{t \alpha}_{t \alpha} =
- \frac{1}{2rB} \ \frac{A^\prime}{A},
$
and
$
R^{r \alpha}_{r \alpha} =
\frac{1}{2rB} \ \frac{B^\prime}{B}, 
$
with the prime denoting differentiation with respect to $r$.

This is the expression for $E_{ab} \ l^a l^b$ 
we obtain assuming the configuration is static.
In the non-static case,
i.e. assuming
$A = A(t, r)$ and $B = B(t, r)$ in (\ref{metric}),
some of the expressions for the components of Riemann tensor
change with respect to the static case
(due to $\dot A \ne 0$ and $\dot B \ne 0$,
where the dot denotes differentiation with respect to $t$),
producing also some components which are no longer vanishing.
These latter are the
$
R^{r \alpha}_{t \alpha} =
\frac{1}{2rB} \ \frac{\dot B}{B}  
$ 
and those related to these by symmetries. 
In the calculations of $E_{tt(m)}$, $E_{rr(m)}$, $E_{rt(m)}$ above, 
the effect of these no-longer-vanishing components
is to give place to additional terms.
As for $E_{rt(m)}$, we have

\begin{eqnarray}
E_{rt(m)} 
&=&
- \frac{1}{2} g_{rt} L_{(m)} + P^{i j}_{k t (m)} {R^k}_{rij} =
P^{i j}_{k t (m)} {R^k}_{rij} 
\nonumber \\
&=&
g_{rr} \ \frac{m}{2^m} \ \delta^{i j c_2 d_2 ... c_m d_m}_{k t a_2 b_2 ... a_m b_m}
\ R^{kr}_{ij} R^{a_2 b_2}_{c_2 d_2} ... R^{a_m b_m}_{c_m d_m}
\nonumber \\
&=&
g_{rr} \ \frac{m}{2^m} 
\ \sum_{\alpha} \delta^{i j c_2 d_2 ... c_m d_m}_{\alpha t a_2 b_2 ... a_m b_m}
\ R^{\alpha r}_{ij} R^{a_2 b_2}_{c_2 d_2} ... R^{a_m b_m}_{c_m d_m}
\nonumber \\
&=&
g_{rr} \ \frac{m}{2^m} \
\Big(
2^3 \sum_{\alpha, \beta} \delta^{\alpha r \beta t c_3 d_3 ... c_m d_m}_{\alpha t \beta r a_3 b_3 ... a_m b_m}
\ R^{\alpha r}_{\alpha r} R^{\beta r}_{\beta t} R^{a_3 b_3}_{c_3 d_3} ... R^{a_m b_m}_{c_m d_m}
+
2^3 \sum_{\alpha, \beta} \delta^{\alpha t \beta r c_3 d_3 ... c_m d_m}_{\alpha t \beta r a_3 b_3 ... a_m b_m}
\ R^{\alpha r}_{\alpha t} R^{\beta r}_{\beta r} R^{a_3 b_3}_{c_3 d_3} ... R^{a_m b_m}_{c_m d_m}
\nonumber \\
& &
+ \
2^m \sum_{\alpha, \alpha_2, \beta_2, ..., \alpha_m, \beta_m \ne}
R^{\alpha r}_{\alpha t} R^{\alpha_2 \beta_2}_{\alpha_2 \beta_2} ... R^{\alpha_m \beta_m}_{\alpha_m \beta_m}
\Big)
\nonumber \\
&=&
g_{rr} \ m
\sum_{\alpha, \alpha_2, \beta_2, ..., \alpha_m, \beta_m \ne}
R^{\alpha r}_{\alpha t} R^{\alpha_2 \beta_2}_{\alpha_2 \beta_2} ... R^{\alpha_m \beta_m}_{\alpha_m \beta_m}
\nonumber \\
&=&
g_{rr} \ (D-2)! \ R^{r \alpha}_{t \alpha} \
\frac{m}{(D-2m-1)!} \ X^{m-1},
\end{eqnarray}
with $X = R^{\alpha \beta}_{\alpha \beta}$ 
given by the same expression (\ref{X})
also for the non-static configuration. 
This gives

\begin{eqnarray}
E_{rt} =
g_{rr} \ (D-2)! \ R^{r \alpha}_{t \alpha} \ Q[X] =  
(D-2)! \ Q[X] \ 
\frac{1}{2rB} \
{\dot B}.
\end{eqnarray}

From this expression,
we have that when, in the spherical
coordinates we consider, $E_{ab}$ is diagonal,
as in vac or cosmovac solutions,
from $E_{rt} = 0$ it follows 
$Q[X] = 0$ or $\dot B = 0$.
This means $\dot B = 0$
(since the $B = \frac{1}{1-{\hat X} r^2}$ from any $\hat X$ with $Q[{\hat X}] = 0$ 
is non-depending on $t$)
and then $R^{r \alpha}_{t \alpha} = 0$, 
even if the configuration we are considering 
is actually not static.
Thus,
for diagonal $E_{ab}$,
the expressions (\ref{p251}-\ref{p253}) remain unchanged 
when going to the non-static case.
Now,
in addition to $R^{\alpha \beta}_{\alpha \beta}$
also the explicit expressions for
$R^{t \alpha}_{t \alpha}$, $R^{r \alpha}_{r \alpha}$
are left unchanged by any $\dot A \ne 0$ (and $\dot B \ne 0$).
The final expression 
of $E_{ab} \ l^a l^b$
in the non-static case
is, then, still equation 
(\ref{p240}) with $A = A(t, r)$ and $B = B(r)$.

We can give
equation (\ref{p240}) 
a slightly different form.
$E_{ab}$ diagonal, implying 
$R^{r \alpha}_{t \alpha} = 0$,
gives $R^r_t =0$.
We have then

\begin{eqnarray}
R_{ab} \ l^a l^b 
&=&
R_{tt} \ l^t l^t + R_{rr} \ l^r l^r 
\nonumber \\
&=&
\mu^2 \ (- g_{tt} g_{rr}) \ (R^r_r - R^t_t)
\nonumber \\
&=&
\mu^2 \ (- g_{tt} g_{rr}) \ (D-2) \ (R^{r \alpha}_{r \alpha} - R^{t \alpha}_{t \alpha})
\end{eqnarray}
and, from the first equality in (\ref{p240}),

\begin{eqnarray}\label{Ricci}
E_{ab} \ l^a l^b 
=
(D-3)! \ Q[X] \ R_{ab} \ l^a l^b,
\end{eqnarray}
with $X = X(r)$ both in the static and non-static case.
The polynomial $Q[X]$,
being from (\ref{Ricci}) the ratio of two scalar quantities,
is itself a scalar.
Looking at its definition (\ref{polQ}),
it can be thougth as invariantly constructed from
its argument $X$, 
which is (eq. (\ref{X})) $R^{\alpha \beta}_{\alpha \beta} (\alpha \ne \beta)$
in the coordinates (\ref{metric}), meant as a scalar.
Also the polynomial $W[X]$, as well as any polynomial in the scalar $X$, 
is thus invariant. 

The results (\ref{p240}) and (\ref{Ricci})
turn out to be entangled with the results \cite{Zeg}.  
$Q$  
coincides with the quantity
denoted as $P^\prime$ in \cite{Zeg}
regarding the dependence 
on their respective arguments
(see eq. (22) there). 
Upon re-transforming back from
the coordinates $(u, v)$
used in \cite{Zeg},
in terms of which
the squared distance (\ref{metric}) is
$
ds^2 = 2 \ {\text e}^{2 \nu(u, v)} \ du dv + {\cal B}^2(u, v) \ d\Omega^2,
$
to the coordinates $(t, r)$ used here,
the argument $Z$ of $P^\prime$,
written there 
in the case of spherical symmetry
as 
$
Z =
\frac{1 \ - \ 2 (\partial_u{\cal B}) (\partial_v{\cal B}) 
{\text e}^{-2\nu}}{{\cal B}^2}
$
is
$
Z = \frac{1}{r^2} (1 - \frac{1}{g_{rr}}) = X,
$
since
$
r = {\cal B}
$
and
$
g_{rr} = \frac{1}{2} 
\frac{{\text e}^{2 \nu}}{(\partial_u{\cal B}) (\partial_v{\cal B})}.
$
$Z$ is thus that same scalar $X$ 
given in (\ref{X})
and 
$
P^\prime\Big[
\frac{1 \ - \ 2 (\partial_u{\cal B}) (\partial_v{\cal B}) 
{\text e}^{-2\nu}}{{\cal B}^2}
\Big]
$
{\it is} actually 
the invariant polynomial $Q[X]$ \cite{Rev}.

Let us consider, as a first use of equations (\ref{p240}) or (\ref{Ricci}),
vac and cosmovac.
From $E_{ab} \ l^a l^b = 0$  in this case, 
we get
$(a)$
$
(AB)^\prime = 0
$
or 
$(b)$ 
$Q[X] = 0$, $A$ generic.
From (\ref{polQ}),
case $b$ can only happen when
at least one of the constants
$c_m$ with $m > 1$ is non-vanishing.
From the coincidence between $P^\prime[Z]$ in \cite{Zeg}
and $Q[X]$ here, this case is that already considered 
as $P^\prime[Z] = 0$ solutions
(class I solutions) in \cite{Zeg},
and a general discussion of their properties 
is given in \cite {CEd}, \cite{MaeB}. 
In our framework, we can notice the following.
In case $b$,
from (\ref{p251})-(\ref{p253})
we have

\begin{eqnarray}\label{Ettb}
E^{(b)}_{tt} = - \frac{1}{2} \ g_{tt} \ (D-1)! \ W[{\hat X}],
\end{eqnarray}

\begin{eqnarray}\label{Errb}
E^{(b)}_{rr} = - \frac{1}{2} \ g_{rr} \ (D-1)! \ W[{\hat X}],
\end{eqnarray}
for any ${\hat X}$ with $Q[{\hat X}] = 0$.
As for $E_{\alpha\alpha}$,
the same algebra which leads to (\ref{p250})-(\ref{p253m}) and (\ref{p251})-(\ref{p253})
gives 

\begin{eqnarray}
E_{\alpha\alpha} 
&=&
\frac{1}{2} \ g_{\alpha\alpha} \ (D-3)! \
\Big\{
2 \ \big[- R^{tr}_{tr} - (D-3) \big(R^{t \beta}_{t \beta} + R^{r \beta}_{r \beta}\big) 
+ (2D - 5) X\big] \ Q[X] 
\nonumber \\ 
& & 
-(D-1) (D-2) \ W[X] - 4 \ \big(X - R^{r \beta}_{r \beta}\big) \big(X - R^{t \beta}_{t \beta}\big) \ Y[X]
\Big\},
\end{eqnarray}
having defined the invariant polynomial

\begin{eqnarray}\label{polY}
Y[X] :=
\sum_{m=1}^{[\frac{D-1}{2}]} \frac{m \ (m-1) \ c_m}{(D-2m-1)!} \ X^{m-2} =
\frac{dQ}{dX} =
\frac{d^2W}{dX^2}.
\end{eqnarray}
In case $b$, this implies

\begin{eqnarray}\label{Eaab}
E^{(b)}_{\alpha\alpha} = - \frac{1}{2} \ g_{\alpha\alpha} \ (D-1)! \ W[{\hat X}],
\end{eqnarray}
since
$Q[{\hat X}] = 0$, 
and when
$B = \frac{1}{1-{\hat X} r^2}$ 
(from any ${\hat X}$ with $Q[{\hat X}] = 0$), 
$R^{r \beta}_{r \beta} = {\hat X}$.   

From (\ref{Ettb}, \ref{Errb}, \ref{Eaab}), 
in case $b$
all what the equations of motion require is 
$W[{\hat X}] = 0$ for vac, and $W[{\hat X}] = \frac{2 \Lambda}{(D-1)!}$ for cosmovac.
A solution, if any,  
has thus $A$ generic.
At the same time, 
the Ricci scalar
$R = \frac{1}{2} \delta^{cd}_{ab} R^{ab}_{cd}$ 
depends on $A$ and its first and second derivatives
and is generically non-vanishing.
This means that different $A$'s generically give spacetimes
with different curvature scalars, 
and thus with genuinely different geometries,
i.e. geometries not recoverable each from the other through
coordinate trasformations.
In case $b$ we have
thus the peculiar circumstances
that 
the solutions for vac or cosmovac,
when they can exist, are under-determined by the equations of motion. 

Case $a$
gives
$AB =$ const,
that is
$AB = f(t)$, with $A = A(t, r)$, $B = B(r)$
and $f$ a function of $t$ alone.
The metric we have is
$ds^2 = -\frac{f(t)}{B(r)} dt^2 + B(r) dr^2 + r^2 d\Omega^2$.
Changing the $t$-coordinate to $\tilde t$ with
$d \tilde t = \sqrt{f} \ dt$, we get
$ds^2 = -\frac{1}{B(r)} dt^2 + B(r) dr^2 + r^2 d\Omega^2$,
thus reducing to $AB =1$ with $A=A(r)$ and $B=B(r)$,
both in the static and non-static cases.
A sensible notion of mass $M$ of gravitating body
can then be given, 
in the form of a generalised 
Misner-Sharp mass \cite{MisB} 
defined in terms of $B(r)$ \cite{MaeA, MaeB, Kun},
and the solutions are parameterized in terms of $M$.

Summing up,
in deriving expressions (\ref{p240}) and (\ref{Ricci}),
and 
considering, as a first example of their use, vac and cosmovac configurations,
we have thus shown the following:

\noindent
{\it Proposition 1.}
Consider
a Lovelock model
and a region with spherically symmetric geometry.
If the generalized Einstein tensor $E_{ab}$ turns out to be diagonal
in the spherical coordinates,
then its two radial null-null components $E_{ab} \ l^a l^b$
(which are equal) can be expressed according to formulae 
(\ref{p240}) and (\ref{Ricci}). \\
\noindent
{\it Proposition 2.}
Consider a Lovelock model
and a region with spherically symmetric geometry.
Any vac or cosmovac solution to the equations of motion
under non-exceptional conditions
(meaning the equations of motion are able to fix the solutions),
can be expressed in the form
$ds^2 = -\frac{1}{B(r)} dt^2 + B(r) dr^2 + r^2 d\Omega^2$,
That is, Birkhoff-like theorem
for Lovelock models.

Another example of use of (\ref{p240}) or (\ref{Ricci}) can be envisaged as follows.
For diagonal, otherwise generic $T_{ab}$ (and thus, in particular, for any
ideal fluid),
from (\ref{fie}) and (\ref{Ricci}) 
we get on-shell

\begin{eqnarray}\label{CC}
T_{ab} \ l^a l^b =
2 \ (D-3)! \ Q[X] \ R_{ab} \ l^a l^b;
\end{eqnarray}
thus, in particular,

\begin{eqnarray}\label{signCC}
\text{sgn}(T_{ab} \ l^a l^b) =
\text{sgn}(Q[X]) \ \text{sgn}(R_{ab} \ l^a l^b),
\end{eqnarray}
with $X$ evaluated for solutions to the equations of motion. 
The NEC
and the NCC 
($T_{ab} \ n^a n^b \geq 0$ and $R_{ab} \ n^a n^b \geq 0$
respectively,
at any point, $\forall n^a$ null) 
in the radial direction
are in general not equivalent in Lovelock models,
and equations (\ref{CC}, \ref{signCC})
trace this fact at any point of our symmetric spacetime
in terms of the invariant polynomial $Q[X]$,
which becomes then a tool to study their relationship
in an invariant manner \cite{comment2}.
The two conditions become equivalent whenever $Q[X] > 0$.
In particular, the NEC and the NCC are equivalent 
in general relativity ($Q[X] = \frac{1}{(D-3)!}$ for it).

As long as we regard
the NEC and the NCC as equivalent expressions 
of some fundamental input required on physical grounds \cite{ParB},
extending however the scope of this 
beyond general relativity to include any metric theory of gravity,
the condition on the sign of $Q[X]$ 
sounds as a physical constraint on viable
Lovelock models and/or on viability of some specific solutions 
to the equations of motion for them.
Namely,
given any solution to the equations of motion 
for some Lovelock model,
the choice made of the constants $c_m$  entering the Lagrangian (\ref{generic_L}),
which defines the model under consideration, 
and the specific solution considered
should guarantee 
that the invariant polynomial $Q[X]$  
in (\ref{polQ}) gives $Q[X] > 0$ at any point of spacetime,
when $X$ is evaluated on the solution \cite{comment3}.

This extending of the NCC and of its equivalence with the NEC 
beyond general relativity
could turn out to be not so ventured after all. 
In view of ($D$-dimensional) Raychaudhuri equation
which applies equally regardless of the model, 
the physical input underlying it
would be to maintain that
`any source matter obeys the NEC' 
and that this is
`its gravity acts 
always focusing' 
in disguise.

Also, 
we can look at this 
from thermodynamics (for thermodynamics
of Lovelock gravity we refer to \cite{ChaA} and references therein).
Indeed, 
$Q[X]$ can be given on-shell the interpretation of ratio of
two (horizon) entropies, and as such it turns out to be quite natural 
to require it to be positive. 

In the l.h.s. of (\ref{CC})
appears, in fact, something
which gives the 
Wald entropy $S_W$ of horizon for the gravity model under consideration.
More precisely,
let us consider
the local frame $\{ x^a\}$ of the element of matter at $P$.
Given a generic null vector $n^a$,
we choose the local frame such that $n^a$
be given by $n^0 = 1$, $n^1 = 1$, with the other components vanishing,
and $x^0(P) = 0$.
Consider an accelerating observer   
along $x^1$ with acceleration $\kappa$,
instantaneously coinciding and at rest with respect
to the element of matter at the time $x^0 = 0$,
and choose the origin of the $x^1$-axis such that
$x^1(P) = \frac{1}{\kappa}$
(let us call it
the accelerating frame 
along $n^a$
of (i.e. associated to) the element of matter).
The Rindler horizon perceived by
the accelerating observer is at
$x^0 \pm x^1 = 0$,
and has, as tangent vector field
to the generators, the field obtained parallel-transporting
$k^a$ in the local Lorentz frame $\{ x^a\}$. 
When the element of matter gets absorbed by the horizon, 
the (semiclassical) variation  $dS_W$ 
of horizon entropy 
in the gravity theory
under consideration
is 
(cf. \cite{PesI})

\begin{eqnarray}
dS_W(P, n^a) 
&=&
\frac{1}{T_H} \int_{-\frac{L}{2}}^{\frac{L}{2}}
T_{ab} \ \xi^a_{(t_R)} n^b A \ d\lambda \nonumber \\
&=&
\frac{1}{T_H} T_{ab} \ n^a n^b \ dV_{prop},
\end{eqnarray}
where $\lambda = x^1 - \frac{1}{\kappa}$ is affine parameter,
$T_H = \frac{\kappa}{2 \pi}$ is the temperature of the horizon,
$\xi^a_{(t_R)}$ is the Killing vector field corresponding
to translations in Rindler time $t_R$,
$dV_{prop}$ is the proper volume of the element of matter
with $L$ proper thickness,
and use of the limit $\xi^a_{(t_R)} \rightarrow \kappa x^1 n^a$ on the
horizon is made. 

As for the r.h.s.,
by the same token
we get that 
the quantity
$
\frac{1}{T_H} (2 \ R_{ab} \ n^a n^b) \ dV_{prop}
$
is the variation of the entropy 
of that same horizon
but for Einstein-Hilbert Lagrangian,
and thus 
is the variation of Bekenstein-Hawking entropy $dS_{BH}(P, n^a)$,
provided that we
consider now that same metric configuration as solution to the equations
of motion of Einstein-Hilbert instead of general Lovelock. 
Indeed,
for $L = L_{(1)}$ in (\ref{generic_L}),
(\ref{fie}) becomes
$
2 R_{ab} - g_{ab} R = T_{ab}.
$
Checking this directly:

\begin{eqnarray}
\frac{1}{T_H} \ (2 R_{ab} \ n^a n^b) \ dV_{prop}
&=&
\frac{1}{T_H} \ 2 \int_{-\frac{L}{2}}^{\frac{L}{2}} 
R_{ab} \ \xi^a_{(t_R)} n^b A \ d\lambda \nonumber  \\
&=&
4 \pi A \int_{-\frac{L}{2}}^{\frac{L}{2}}
R_{ab} \ n^a n^b \ \Big(\lambda + \frac{1}{\kappa}\Big) \ d\lambda  \nonumber \\
&=&
4 \pi A \int_{-\frac{L}{2}}^{\frac{L}{2}}
\Big(-\frac{d\theta}{d\lambda}\Big) 
\ \Big(\lambda + \frac{1}{\kappa}\Big) \ d\lambda  \nonumber \\
&=&
4 \pi A
\ \bigg\{ 
\Big[-\theta \ \ \Big(\lambda + \frac{1}{\kappa}\Big)\Big]_{-\frac{L}{2}}^{\frac{L}{2}}
+ 
\int_{-\frac{L}{2}}^{\frac{L}{2}} \theta \ d\lambda
\bigg\}  \nonumber \\
&=&
4 \pi A \int_{-\frac{L}{2}}^{\frac{L}{2}} \theta \ d\lambda \nonumber \\
&=&
4 \pi A \ \frac{dA}{A} \nonumber \\
&=&
4 \pi \ dA 
=
dS_{BH}(P, n^a)
\end{eqnarray}
(with the normalization
$
L = \frac{1}{16 \pi} R = \frac{1}{16 \pi} L_{(1)},
$
we get
$
\frac{1}{16 \pi} (2 R_{ab} - g_{ab} R) = T_{ab}
$
and
$
\frac{1}{16 \pi}
\frac{1}{T_H} (2 R_{ab} \ n^a n^b) \ dV_{prop} =
\frac{dA}{4}),
$
where
$
\theta = \frac{d\ln{A}}{d\lambda}
$
is the expansion of the null congruence
generating the horizon. 
Here,
use has been made of
$
\theta\big(-\frac{L}{2}\big) = 0 = \theta\big(\frac{L}{2}\big)
$
to express initial and final stationarity of the horizon,
and of the approximation
$
\frac{d\theta}{d\lambda} = - R_{ab} \ n^a n^b
$
for $\theta$ small.

We can thus recast (\ref{CC})
in the form

\begin{eqnarray}
dS_W(P, l^a) = (D-3)! \ Q[X] \ dS_{BH}(P, l^a)
\end{eqnarray}
or

\begin{eqnarray}
Q[X] = \frac{1}{(D-3)!} \ \frac{dS_W(P, l^a)}{dS_{BH}(P, l^a)}.
\end{eqnarray}
This shows that the invariant polynomial $Q[X]$ has also
the following thermodynamical meaning:
Considering
some spherically symmetric configuration,
solution to the equation of motions of some Lovelock gravity model,
and an accelerating frame relative to matter
along the radial direction
with some element of matter getting absorbed
by the Rindler horizon,
$Q[X]$
is \big($\frac{1}{(D-3)!}$ times\big) the ratio
of
the entropy variation of the horizon
in the Lovelock gravity model under consideration
to
the variation of Bekenstein-Hawking entropy
one would obtain
when considering the same configuration
as solution of Einstein's equations. 

In this perspective,
the physical input underlying
the extending of the NCC and of its equivalence with the NEC 
beyond general relativity
could be imagined as follows:
Any particle which gets absorbed by a Rindler horizon
gives a positive (semiclassical) variation of the entropy
of the horizon, whichever is the gravity metric theory actually chosen by Nature,
theory which prescribes the way to assign entropy to the horizon.
On this basis,
the invariant polynomial $Q[X](P)$
could turn out to be the form 
that a much more general quantity $q(P, n^a)$ takes
when evaluated for Lovelock theories and in the radial
direction of spherical-symmetry configurations,
being $q(P, n^a)$ defined as

\begin{eqnarray}
q(P, n^a) := \frac{1}{(D-3)!} \ \frac{dS_W(P, n^a)}{dS_{BH}(P, n^a)}
\end{eqnarray}
in any diff-invariant theory of gravity,
for any absorption process at $P$
by a Rindler horizon 
with whichever tangent vector $n^a$ to the horizon generators,
and for generic configurations.

\section{Comments and conclusions}

We have provided, in equations (\ref{p240}) and (\ref{Ricci}),
expressions for the radial null-null components $E_{ab} \ l^a l^b$
of generalized Einstein tensor for Lovelock models
for diagonal $E_{ab}$.
And shown they can be used in deriving 
the Birkhoff-like theorem for `generic' such models,
as well as 
in discriminating among the models and/or among
specific solutions to them  
in terms of the physical input provided by
the null energy and null convergence conditions.

The validity of Birkhoff-like theorem means
that any spherically symmetric vac or cosmovac solution
to a generic Lovelock model is static, and its metric can be put in the form
$g_{tt} g_{rr} = -1$.
A general condition, given in \cite{JacF},
for a sperically symmetric static metric to have this form,
is as mentioned the vanishing of 
the radial null-null components of Ricci tensor $R_{ab} \ l^a l^b$.
We expect thus
that, for spherically symmetric vac or cosmovac solutions
to generic Lovelock, $R_{ab} \ l^a l^b =0$.
And this is precisely what happens,
since,
from expression (\ref{Ricci}), 
$E_{ab} \ l^a l^b$ vanishes when
$R_{ab} \ l^a l^b$ vanishes.

In the approach we have described,
the validity of Birkhoff-like theorem
is read in the expression
for $E_{ab} \ l^a l^b$:
in that $B=B(r)$
and in that the vanishing of $E_{ab} \ l^a l^b$
means $(AB)^\prime = 0$.
Even a `minimal' departure from LL gives troubles.
Taking , for example, ${\tilde L} = f(R) = R^2$,
thus restricting consideration to the effects
of this particular element of
the Gauss-Bonnet term alone, 
we have
${\tilde L} = \frac{1}{2} \ \delta^{cd}_{ab} R \ R^{ab}_{cd}
= \frac{1}{2} \frac{\partial {\tilde L}}{\partial R^{ab}_{cd}} \ R^{ab}_{cd}
= \frac{1}{2} P^{cd}_{ab} \ R^{ab}_{cd}$,
being $\frac{\partial {\tilde L}}{\partial R^{ab}_{cd}} = \delta^{cd}_{ab} R$.
${\tilde L}$ is in the form
${\tilde L} = Q^{cd}_{ab} R^{ab}_{cd}$ with
$Q^{cd}_{ab}$ still being polynomial 
(linear indeed) in the components of Riemann and 
having the same symmetries of Riemann
as in LL eq. (\ref{LLL});
the only change is the relaxing
of the condition (\ref{div}) on the divergence of $Q^{abcd}$.
The expression for
${\tilde E}_{ab} \ l^a l^b$ one obtains 
following the lines here,
replacing eq. (\ref{p240}),
contains indeed 4-th order derivatives of the metric,
analogously to what happens for the equations of motion
(an account of Birkhoff-like theorems in $f(R)$ theories can be found in \cite{NGD};
an investigation of the conditions which theories with equations of motion
of order larger than 2 in the derivatives of the metric 
should obey for Birkhoff-like theorem to hold, is in \cite{OR_A, OR_B}).



\begin{thebibliography}{00}
\bibitem{comment4}
The equations of motion are quasi-linear differential equations.
Requiring this, is what singles out the Lovelock terms uniquely. 
Equivalently, one requires 
the generalized Einstein tensor
is coming from
a 4th rank tensor homogeneous in the components of Riemann, 
the trace of the Bianchi derivative of which 
be divergence-free \cite{DadA, Kas, CamB}.
\bibitem{Rev}
Equations (20) and (21) of \cite{Zeg} turn out
to be strongly tied to equations (\ref{p240}) or (\ref{Ricci})
here.
%
Writing eqs. (20) and (21) of \cite{Zeg} as
$U = 0$ and $V = 0$ 
(thus with
$
U = P^\prime[Z] 
\ [\partial^2_u{\cal B} - 2 (\partial_u{\cal B}) (\partial_u\nu)]
$
and
$
V = P^\prime[Z]
\ [\partial^2_v{\cal B} - 2 (\partial_v{\cal B}) (\partial_v\nu)]
$),
we find
$
2 r E_{ab} \ l^a l^b = 
- \mu^2 \ (D-2)! \ A \
\Big[\frac{1}{(\partial_u{\cal B})^2} U 
+ \frac{1}{(\partial_v{\cal B})^2} V\Big]
= 
- (D-2)! \
[(l_+^u)^2 \ U + (l_-^v)^2 \ V],
$
with $l_+^a$ ($l_-^a$) the radial null vectors
with sign $+$ ($-$) in (\ref{radialnull}) above,
with components
$l_+^u = \mu \ \frac{1}{\partial_u{\cal B}} \sqrt{A}, \ 
l_+^v = 0, \ 
l_+^\alpha = 0$
($l_-^u = 0, \ 
l_-^v = - \mu \ \frac{1}{\partial_v{\cal B}} \sqrt{A}, \ 
l_-^\alpha = 0$)
in $(u, v)$ coordinates.
\bibitem{comment2}
For the case 
of Einstein-Gauss-Bonnet gravity,
a formula aiming too at relating 
NEC and NCC in the radial direction 
has been given
in \cite{Noz} eq. (3.14).
\end{thebibliography}
\end{document}